\documentclass[11pt]{article}
\usepackage{graphicx}
\usepackage[utf8]{inputenc}
\usepackage[T1]{fontenc}
\usepackage{lmodern}

\usepackage{url}
\usepackage{authblk} 

\title{Permanent Data Encoding (PDE): A Visual Language for Semantic Compression and Knowledge Preservation in 3-Character Units}

\author{
Yoshiharu Tsuyuki\thanks{Departments of Oral and Maxillofacial Surgery, Yokohama City University \& Matsumoto Dental University. Corresponding author: \texttt{fwkk8987@nifty.com}},\ 
Xianqi Li\thanks{Department of Oral and Maxillofacial Surgery, Matsumoto Dental University},\ 
Yuji Kurihara\thanks{Department of Oral and Maxillofacial Surgery, Matsumoto Dental University},\ 
Kenji Mitsudo\thanks{Department of Oral and Maxillofacial Surgery, Yokohama City University}
}

\date{\today} 

\begin{document}

\maketitle

\begin{abstract}
\noindent
Permanent Data Encoding (PDE) is a visual language framework designed for long-term, human-readable, and electrically independent knowledge preservation. By encoding semantic content into compact, 2–3 character alphanumeric codes—paired with public dictionaries and rule-based expansion structures—PDE enables information to be visually interpreted and logically reconstructed without reliance on digital systems.

\noindent
Unlike QR codes or binary data, which require machines to decode, PDE offers a semiotically transparent and structurally self-contained method of encoding meaning. It is designed to withstand technological obsolescence, media degradation, and sociocultural disruption. The core components of PDE include a syntax system, a modular vocabulary structure, and a blockchain-based dictionary to ensure semantic immutability and global verifiability.

\noindent
This paper outlines the design principles behind PDE, compares it with existing symbolic and linguistic systems, and introduces its syntactic and semantic encoding architecture. Practical use cases are examined in the contexts of disaster resilience, multilingual education, and human-AI interaction. Limitations and open challenges are also discussed, including dictionary management, cross-linguistic scalability, and symbolic compression boundaries.

\noindent
PDE is ultimately proposed as a visual-semantic infrastructure—a hybrid between language and code—that offers a new approach to storing and transmitting knowledge across generations. It seeks not only to preserve data, but to preserve meaning itself.
\end{abstract}

\section{Introduction}
In today's digital society, the generation and distribution of information are advancing at an unprecedented pace. However, a critical question arises: are we truly able to preserve information over the long term? While storage media such as HDDs, SSDs, and cloud platforms offer decades of reliability, their longevity over a century or a millennium remains uncertain. Moreover, these systems are deeply dependent on electricity, hardware, operating systems, and other infrastructures—making them fragile in the face of disasters, wars, or technological obsolescence.

\noindent
Modern information storage systems are typically not designed for human interpretation without the aid of machines. For example, QR codes and binary data cannot be understood directly by the human eye; they require specific devices and decoding algorithms. This dependency raises the possibility that much of today's recorded information may become inaccessible "black boxes" to future generations.

\noindent
To address these challenges, this paper proposes Permanent Data Encoding (PDE)—a novel visual language system for semantic compression and long-term, human-readable information preservation. PDE represents meaning using compact, visually recognizable 2–3 character alphanumeric codes, paired with a publicly defined dictionary and expansion rules. This design enables humans to decode and reconstruct information with the naked eye and basic reasoning, without reliance on technology.

\noindent
Crucially, PDE dictionaries are maintained using blockchain technology to ensure immutability, decentralization, and semantic consistency. As a result, PDE functions both as a paper-based archival system and a digitally processable encoding format. Its hybrid nature allows for resilience across time, culture, and infrastructure—positioning PDE not merely as a data compression method, but as a universal, visual-semantic language capable of preserving human knowledge across generations.

\noindent
The discussion begins by reviewing the limitations of existing storage methods and symbolic systems. It then introduces the core structure and grammar of PDE, its blockchain-based dictionary architecture, and representative applications in areas such as disaster resilience, education, and post-civilizational recovery.

\noindent
The paper concludes with an exploration of PDE's philosophical, linguistic, and technological significance, along with future directions for international development.

\section{Design Principles and Conceptual Background}

This section outlines the theoretical and technological foundations underlying Permanent Data Encoding (PDE). The design of PDE draws from four primary domains: (1) visual symbolic systems, (2) information compression and recovery methods, (3) long-term digital preservation strategies, and (4) blockchain-based integrity management.

\subsection{Visual Symbol Systems and Language Structures}
PDE builds upon the idea that visual symbols can convey meaning across linguistic and cultural boundaries. Historical examples include Otto Neurath's ISOTYPE (International System of Typographic Picture Education) and modern pictograms used in public signage. While these systems allow for intuitive recognition, they often lack the syntactic complexity to represent nuanced meaning or recreate full natural language expressions.

\noindent
Constructed languages such as Esperanto aim for semantic consistency and neutrality but require substantial learning and are not easily interpreted visually. PDE situates itself between these paradigms—combining visual recognizability with a semantic compositional structure, enabling both readability and expressive depth.

\subsection{Information Compression and Recovery Techniques}
Traditional compression methods such as Huffman coding, run-length encoding, and tokenization in natural language processing (NLP) emphasize computational efficiency and digital compactness. However, they are not intended to be human-readable or interpretable without machine assistance.

\noindent
PDE, by contrast, employs fixed-length alphanumeric codes (e.g., G2, s01, p02) that correspond to discrete semantic units. These codes are organized using a publicly defined dictionary, enabling human users to reconstruct the original meaning without digital devices. This balance of semantic compression and cultural translatability forms the core philosophy of PDE.

\subsection{Long-Term Preservation and Media Independence}
Digital media such as CD-Rs, hard drives, and cloud platforms have finite lifespans and rely on compatible devices, file formats, and energy sources. These dependencies make digital records vulnerable to loss during disasters, wars, or technological obsolescence.

\noindent
PDE addresses these risks by enabling human-readable, paper-compatible data encoding that can be interpreted without electricity or machines. Its visual code structure is resilient to degradation, printing artifacts, and media aging—ensuring accessibility even in post-technological contexts.

\subsection{Blockchain for Semantic Consistency and Trust}
To maintain consistency in code-to-meaning mappings, PDE employs blockchain technology as a decentralized, tamper-resistant ledger for its semantic dictionary. These components—listed in Table~\ref{tab:blockchain-benefits}---ensure immutability, authorship clarity, and semantic transparency across versions and domains.

\begin{table}[h]
\caption{Characteristics and Benefits of Using Blockchain in the PDE Dictionary}
\label{tab:blockchain-benefits}
\centering
\resizebox{\textwidth}{!}{%
\begin{tabular}{ll}
\hline
\textbf{Characteristic} & \textbf{Benefit for the PDE Dictionary} \\
\hline
Immutability & Registered definitions cannot be altered \\
Decentralization & The dictionary is not managed by a central authority \\
Verifiability & Each code definition can be verified by third parties \\
Traceability & All history of additions and modifications is recorded \\
\hline
\end{tabular}
}
\end{table}

\section{Code Taxonomy and Lexical Design}

PDE codes are structurally designed for human readability, syntactic clarity, and scalability. They include short, fixed-length alphanumeric codes—mainly 2-character syntax markers and 3-character semantic units.

\noindent
These codes are grouped into short, mid, and long ranges depending on functional requirements, as summarized in Table 2.
\begin{table}[h]
\caption{Code Lengths and Their Functional Classifications}
\label{tab:code-lengths}
\centering
\resizebox{\textwidth}{!}{%
\begin{tabular}{|l|l|l|}
\hline
\textbf{Code Length} & \textbf{Use} & \textbf{Functional Layer \& Properties} \\
\hline
2 characters & Syntax control code (grammar elements) & 
\textbf{Category level}: frequent, short, highly distinguishable \\
\hline
3 characters & Semantic code (words/events) & 
\textbf{Concept level}: diverse content, everyday language \\
\hline
5 characters & Extended compound terms & 
\textbf{Detailed level}: expert or unknown term encoding \\
\hline
\end{tabular}
}
\end{table}
These codes are defined by a declarative dictionary and can be processed by both human and machine agents.

\subsection{Code Structure and Layered Design}
PDE adopts a three-tier architecture:

\noindent
\textbf{Syntax Control Codes (2 characters)} – denote grammatical roles, such as subject markers (j1), tense (tn), or negation (nG).

\medskip

\noindent
\textbf{Vocabulary Codes (3 characters)} – represent semantic content, e.g., \texttt{p02} for ``woman'', \texttt{a11} for ``sit''.

\medskip

\noindent
\textbf{Extended Compound Codes (5+ characters)} – allow encoding of domain-specific technical terms by combining control and vocabulary units (as shown in Figure~\ref{fig:pde-hierarchy}).

\begin{figure}[h]
\centering
\includegraphics[width=0.6\textwidth]{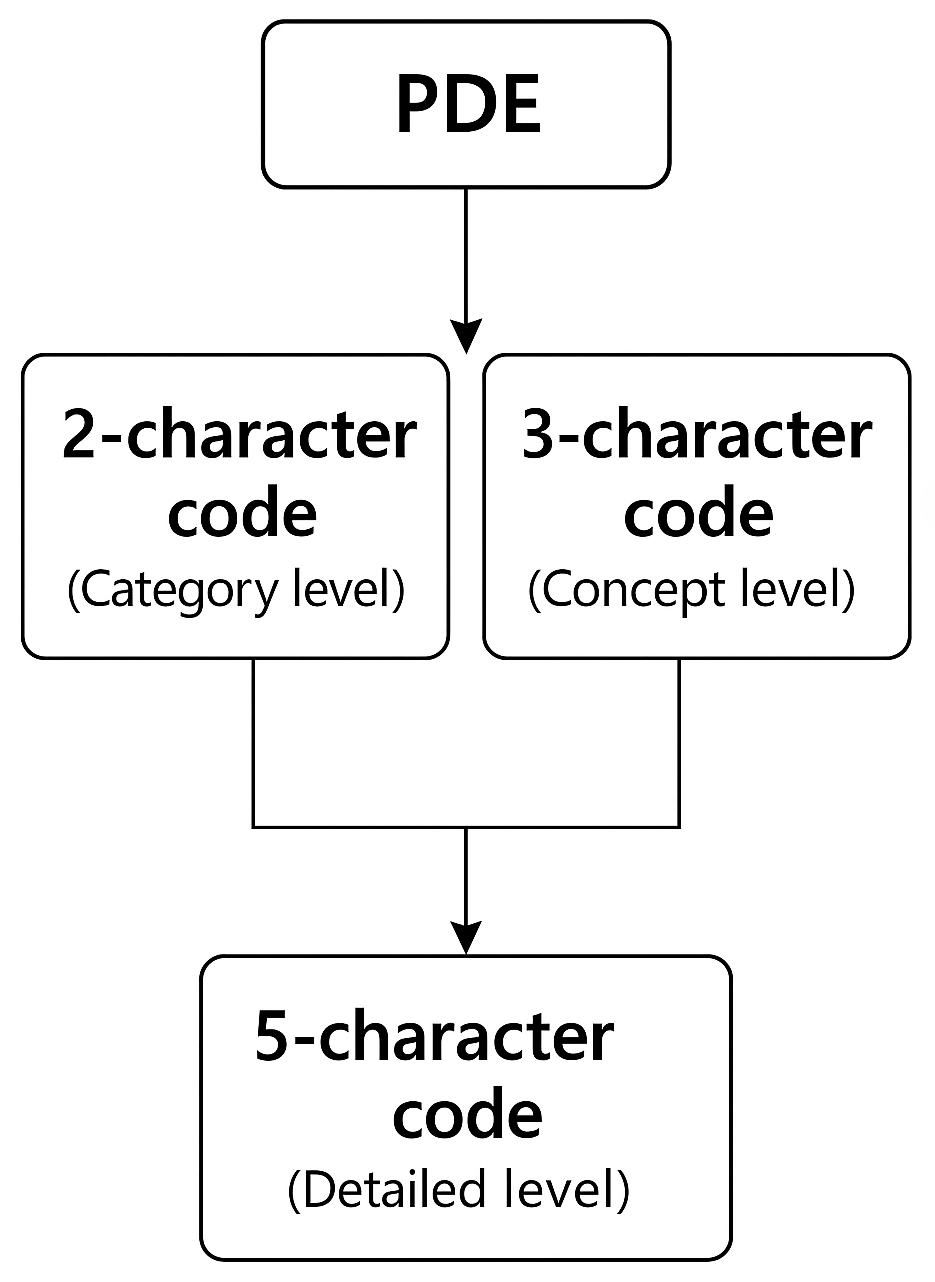}
\label{fig:pde-hierarchy}
\caption{Hierarchical structure of PDE codes}
\label{fig:pde-hierarchy}
\end{figure}

\noindent
Each PDE sentence begins with a declaration prefix (PDE) and ends with a single period (.) to denote the completion of a semantic unit. When multiple sentences form a cohesive block or message, a double period (. .) is appended to mark the end of the entire block. In both cases, the period(s) are written immediately after the final code without a space, ensuring clarity and consistency. This structural rule supports automatic parsing, accurate segmentation, and context-aware restoration of meaning.

\noindent
The core of PDE is built upon a layered code system: each code represents a specific semantic element (e.g., actor, action, time, context), and the combination of these codes forms a minimal yet expressive sentence. By clearly separating syntactic structure from semantic content, PDE ensures that meaning remains both compressible and reconstructable.

\noindent
Below are two illustrative examples:

\medskip

\noindent
\textbf{Example 1:}
\begin{verbatim}
==== Data Codes ====  
PDE p02 j1 tn a11.  
==== Decoded ==== 
The woman sat.
\end{verbatim}

\medskip

\noindent
\textbf{Example 2:}
\begin{verbatim}
==== Data Codes ==== 
PDE p01 j1 a11. p02 j1 tn a11..  
==== Decoded ====
The man sits. The woman sat.
\end{verbatim}

\medskip

\noindent
These examples demonstrate how PDE maps abstract meanings into compact, interpretable syntax units. Each code corresponds to a distinct semantic feature—such as subject (p01/p02), verb (j1), aspect (a11), and tense (tn)—and these features are consistently ordered and delimited.

\noindent
By mapping each code to a specific semantic element, PDE maintains a consistent syntactic structure that is independent of natural language or domain.  
This separation of syntax and semantics promotes clarity, reusability, and extensibility across diverse contexts and applications.

\subsubsection{Embeddability and Modularity}
A PDE block is fully self-contained and delimited by a unique prefix (`PDE`) and a double-period suffix (`. .`). This modular structure allows PDE sentences to be seamlessly embedded into various external systems—such as source code comments, HTML/XML markup, JSON metadata fields, or printed documentation—without requiring any external parser or interpreter.

\noindent
Because PDE blocks are syntactically independent, they can be detected, extracted, and interpreted with simple pattern-matching techniques, even within heterogeneous environments. This design significantly extends the applicability of PDE to cross-domain use cases including medical records, cultural preservation, digital contracts, and long-term physical archiving.

\noindent
In this sense, the PDE block acts not merely as a line of data, but as a portable semantic unit: a piece of meaning that survives across formats, technologies, and time.

\subsection{Semantic Categories and Prefix Convention}
Vocabulary codes are classified using semantic prefixes that correspond to broad categories such as person, color, shape, and emotion.
\noindent
These prefixes enhance readability, dictionary organization, and semantic predictability, especially in multilingual or domain-specific applications. Table~\ref{tab:semantic-prefixes} provides representative examples.

\begin{table}[h]
\caption{Semantic Categories and Their Prefixes in PDE}
\label{tab:semantic-prefixes}
\centering
\resizebox{\textwidth}{!}{%
\begin{tabular}{|l|l|l|}
\hline
\textbf{Category} & \textbf{Prefix Example} & \textbf{Meaning Example / Description} \\
\hline
Person      & pXX  & p02 = female \\
Color       & CXX  & C03 = white \\
Shape       & sXX  & s01 = square \\
Background  & bGX  & bG1 = sea \\
Time        & tXX  & t30 = evening \\
Action      & aXX  & a11 = sit \\
Emotion     & e0X  & e02 = sadness \\
\hline
\end{tabular}
}
\end{table}
\noindent
Vocabulary codes are classified using semantic prefixes that correspond to broad categories such as:
These prefixes enhance readability, dictionary organization, and semantic predictability, especially in multilingual or domain-specific applications. Table 3 provides representative examples.

\subsection{Visual Legibility and Character Constraints}
PDE prioritizes visual clarity, especially in analog media (e.g., printed paper). To minimize misreading, visually ambiguous characters are excluded or replaced. Specifically, the following characters are restricted:

\noindent
This visual filtering draws from best practices in software interfaces, labeling systems, and industrial design—yet PDE integrates these constraints at the structural level of the language itself, ensuring long-term human interpretability.

\subsection{Vocabulary Scalability and Code Economy}
Rather than attempting to replicate a full natural language dictionary, PDE focuses on semantic minimalism and recombinability. Linguistic studies show that:

\begin{itemize}
  \item Daily communication uses \textasciitilde3,000–5,000 core words.
  \item Even specialized professionals rarely use more than 20,000–30,000 active terms.
\end{itemize}

\noindent
PDE targets a core vocabulary size of 10,000–20,000 codes, which can express richer concepts via:

\medskip

\noindent
\textbf{Grouping Example:}
\begin{verbatim}
G2 a12 b34
→ Compound meaning from two adjacent codes
\end{verbatim}

\medskip

\noindent
\textbf{Extension Example:}
\begin{verbatim}
x2 ab C15
→ Integrated domain-specific compound (abC15)
\end{verbatim}

\medskip

\noindent
This economy enables global standardization, easier dictionary maintenance, and code memorability.

\subsection{Design Philosophy: Language as Visual Code}
Unlike systems that merely avoid ambiguous characters as a secondary consideration, PDE enforces design-level exclusion of visually confusable symbols as a core structural policy. This approach aligns with best practices found in:

\begin{itemize}
  \item Programmer-optimized fonts
  \item Secure identification systems
  \item Industrial labeling standards
\end{itemize}

\begin{table}[h]
\caption{Characters excluded or substituted in PDE due to visual similarity issues}
\label{tab:character-substitution}
\centering
\resizebox{\textwidth}{!}{%
\begin{tabular}{|l|l|l|}
\hline
\textbf{Excluded Character} & \textbf{Confusing With} & \textbf{Substitution in PDE} \\
\hline
I (uppercase i)       & 1, l (lowercase L)       & \texttt{i} (lowercase i) \\
O (uppercase o)       & 0                        & \texttt{o} (lowercase o) \\
l (lowercase L)       & 1, I                     & \texttt{L} (uppercase L) \\
B (uppercase B)       & 8                        & \texttt{b} (lowercase b) \\
S (uppercase S)       & 5                        & \texttt{s} (lowercase s) \\
Z (uppercase Z)       & 2                        & \texttt{z} (lowercase z) \\
g (lowercase g)       & 9, q                     & \texttt{G} (uppercase G) \\
U (uppercase U)       & V, Y                     & \texttt{u} (lowercase u) \\
c (lowercase c)       & e, o                     & \texttt{C} (uppercase C) \\
\hline
\end{tabular}
}
\end{table}
\noindent
However, PDE extends beyond these precedents by establishing legibility as a foundational design principle of the language itself. Each code is deliberately crafted to withstand media degradation, copying artifacts, and non-digital environments, thereby ensuring that human readers—even without digital assistance—can confidently interpret encoded content.

\section{Syntax Control Codes and Symbolic Grammar}

While PDE codes can represent individual semantic units, complex meaning construction requires an additional layer of syntactic control. PDE introduces a suite of 2-character control codes that govern the grammatical structure, logical relationships, and expressive nuances of code sequences. This symbolic grammar enables PDE to support structured, language-independent information encoding.

\subsection{Grouping and Unit Designation (rx)}
The \texttt{rx} code defines the number of subsequent tokens to be interpreted as a single semantic unit.

\medskip  

\noindent
\textbf{Examples:}
\begin{verbatim}
r3 C01 s01 p02
→ Interpreted as "a woman with a white square"
→ r3 signals that the next three codes form a compound noun phrase.
\end{verbatim}

\subsection{Modifier Grouping (Gx)}
\texttt{Gx} codes allow the bundling of modifiers and nouns to form natural descriptive units.

\medskip  

\noindent
\textbf{Examples:}
\begin{verbatim}
G2 C02 y01
→ "blue sky"
→ C02 = blue, y01 = sky
\end{verbatim}

\subsection{Code Extension (xX)}
Compound technical terms or domain-specific phrases are encoded via extension codes.

\medskip  

\noindent
\textbf{Examples:}
\begin{verbatim}
x2 ab C15
→ abC15 ("fast-breeder reactor")

x3 C02 y01 z07
→ C02y01z07 (a 7-character technical code)
\end{verbatim}

\noindent
These allow scalable vocabulary without overloading the core dictionary.

\subsection{Semantic Relationship Codes (ux)}
These codes define spatial, temporal, or logical relationships.

\medskip  

\noindent
\textbf{Examples:}
\begin{verbatim}
u1 s01 → "on the square"
u2 t20 → "after sunset"
\end{verbatim}

\noindent
This abstraction reduces the need for language-specific prepositions or inflections.

\subsection{Particle-like Grammatical Roles}
PDE uses symbolic codes to represent grammatical particles such as subjects and objects:

\begin{verbatim}
j1 = marks the subject
o1 = marks the object
\end{verbatim}

\noindent
\textbf{Examples:}
\begin{verbatim}
p01 j1 a11     → "The man sits"
a03 o1 o15     → "Throw the ball"
\end{verbatim}

\noindent
This structure supports both SVO (Subject-Verb-Object) and SOV syntax models, enabling adaptation to multiple natural languages.

\subsection{Tense, Negation, and Comparison}
Temporal states, negations, and comparative expressions are encoded through designated control codes:

\medskip  

\noindent
\textbf{Examples:}
\begin{verbatim}
Past tense:     tn       →  p02 tn a11     → "The woman sat"
Negation:       nG       →  nG a03         → "Not walk"
Comparison:     Cp       →  C01 Cp C02     → "Red vs. blue"
\end{verbatim}

\noindent
These modifiers attach to verbs or nouns and influence sentence-level interpretation.

\subsection{Stylistic and Affective Adjustments}
Emotive tone, formality, and referential expressions can be symbolically encoded:

\medskip  

\noindent
\textbf{Examples:}
\begin{verbatim}
i3 e02         → "Very sad"
st             → Indicates colloquial speech style
d1 s01         → "This square"
\end{verbatim}

\noindent
This design supports style-level control for storytelling, user interfaces, or emotionally tagged communication.

\section{Expansion Rules and Basic Syntax Patterns}

\begin{figure}[h]
\centering
\includegraphics[width=0.5\textwidth]{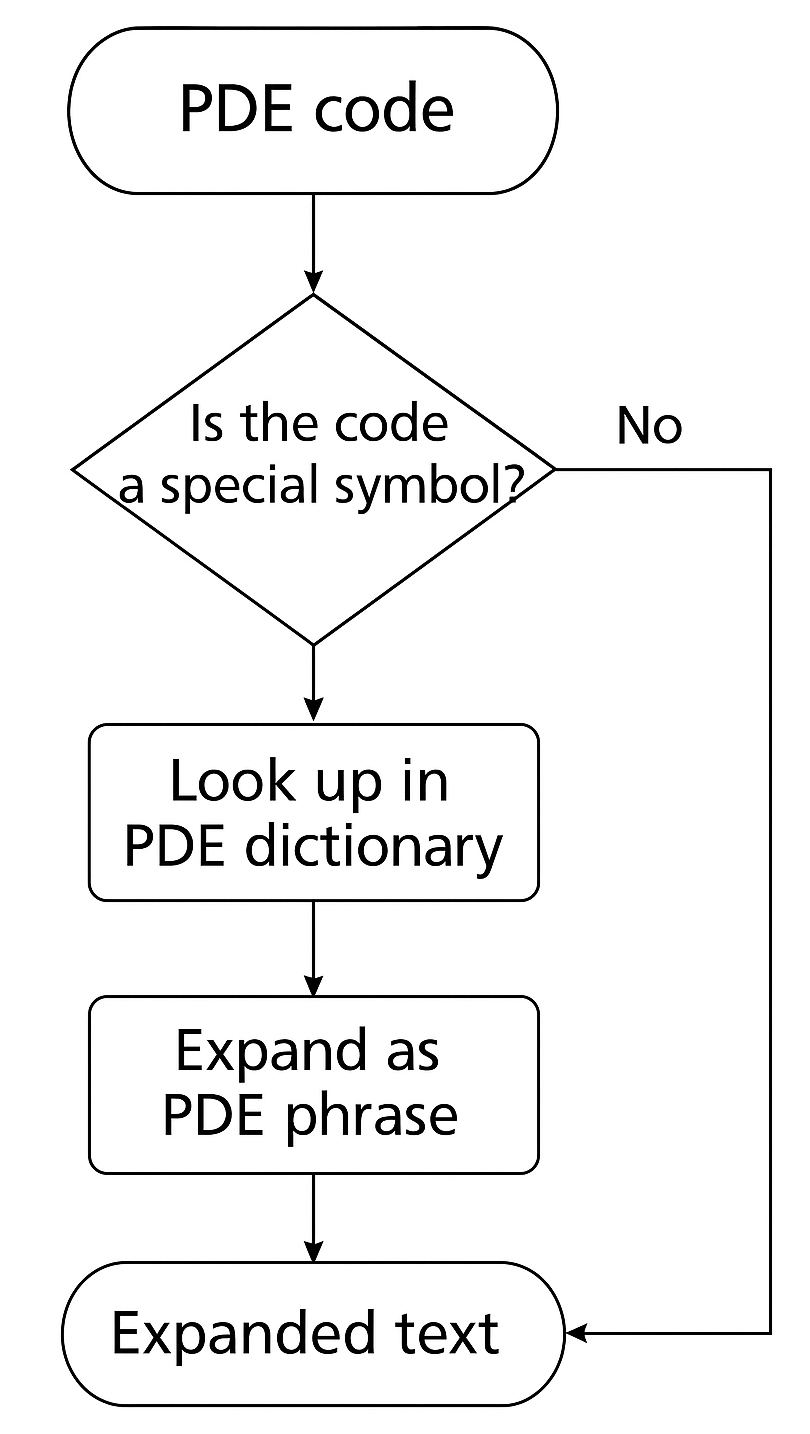}
\label{fig:pde-decoding}
\caption{Syntax rules for expanding PDE}
\label{fig:expansion-flow}
\end{figure}

PDE is not merely a set of codes, but a structured language system designed to expand into natural sentences or semantic units through well-defined syntactic rules. These rules govern how symbolic units are grouped, sequenced, and transformed into human-readable forms. Figure~\ref{fig:expansion-flow} illustrates the overall process of this symbolic expansion.

\subsection{Basic Expansion Template}
Each PDE data line consists of a sequence of codes representing meaning, structure, and style. These are expanded according to semantic and syntactic logic.

\medskip

\medskip  

\noindent
\textbf{Examples:}
\begin{verbatim}
==== Data Codes ====
PDE p02 j1 a11 u1 s01 bG1 t30 e02.
==== Decoded ====
The woman sits on a square. The background is the sea. 
Time: evening. Emotion: sadness.
\end{verbatim}

\medskip
\noindent
This template supports compact, visually organized semantic expression with optional attributes such as location, background, time, and emotion.

\subsection{Sentence Pattern Templates}
Typical expansion follows natural sentence structures:

\begin{verbatim}
{Subject} {verb} {object}
{Subject} {is/does} {action} at {place}
Background is {X}. Time is {Y}. Emotion: {Z}
\end{verbatim}

\noindent
The code order reflects logical grouping, while the expansion rule adjusts word order for natural flow depending on the target language.

\subsection{Effects of Structural Codes}
Grouping Codes (Gx, rx) allow compound noun or verb phrases to be reconstructed correctly. Particle Codes (j1, o1) clarify sentence roles. Tense (tn) and negation (nG) modify verb semantics. These structural codes act as meta-tags, enabling flexible sentence assembly across grammatical systems.

\medskip  

\noindent
\textbf{Examples:}
\begin{verbatim}
PDE u1 G2 C03 s01 p02 j1 nG a11 bG1 e02.
Expanded:
The woman does not sit on a white square. 
Background: sea. Emotion: sadness.
\end{verbatim}

\noindent
Even with minimal codes, PDE supports nuanced expressions through symbolic construction. Natural language rendering is governed by a combination of code hierarchy and rule-based sequencing.

\subsection{Numeric Values and Expressions}
In PDE, numeric information is expressed using number tokens prefixed with the \# symbol (e.g., \#12). There is no limitation on the number of digits. Basic arithmetic operations such as addition, subtraction, multiplication, and division are represented using internationally recognized ASCII symbols (+, -, *, /), ensuring both human readability and machine interpretability.

\noindent
For more advanced mathematical expressions, PDE adopts conventional notation commonly used in English-based technical contexts, including the following.
\begin{itemize}
  \item \verb|^| for exponentiation
  \item \verb|sqrt()| for square roots
  \item \verb|log()| for logarithmic functions
\end{itemize}

\noindent
Parentheses \verb|()| are used to clarify the order of operations, providing visual structure for nested or complex expressions. 

\begin{table}[h]
\caption{Examples of Arithmetic Expressions in PDE}
\label{tab:arithmetic-expressions}
\centering
\begin{tabular}{|l|l|}
\hline
\textbf{PDE Expression} & \textbf{Interpreted Meaning} \\
\hline
\verb|2 + 3| & 2 plus 3 \\
\hline
\verb|10 / 2| & 10 divided by 2 \\
\hline
\verb|5 * (3 + 4)| & 5 multiplied by (3 plus 4) \\
\hline
\verb|2 ^ 3| & 2 raised to the power of 3 \\
\hline
\verb|sqrt(16)| & square root of 16 \\
\hline
\verb|log(100)| & logarithm of 100 \\
\hline
\end{tabular}
\end{table}

\noindent
This syntactic approach allows even mathematically rich content to be rendered in a format that is both easily parsed by machines and intuitively readable by humans on paper. 

\section{Blockchain-Based Dictionary Management System}
At the heart of PDE is a publicly defined dictionary that maps compact codes to semantic meanings. To preserve the integrity, consistency, and historical traceability of these mappings, PDE adopts a blockchain-based dictionary management system. This section outlines the rationale, structure, and operational protocol of this semantic registry.

\subsection{Why Blockchain?}

Traditional dictionary systems rely on centralized servers or local files, which are vulnerable to:

\begin{itemize}
  \item Unauthorized edits or deletions
  \item Version inconsistency and data corruption
  \item Ambiguity in definition ownership and authorship
  \item Loss of historical records and auditability
\end{itemize}

\noindent
Blockchain technology offers a solution by providing a tamper-resistant, decentralized ledger that can ensure transparency and immutability. In PDE, this infrastructure underpins semantic trust.

\subsection{Ensuring Semantic Integrity}

Each PDE code (e.g., p02, bG1) is tied to a definition that includes:
\begin{itemize}
  \item Meaning (e.g., "woman", "sea")
  \item Semantic category (e.g., person, background)
  \item Author and timestamp
  \item Hashed definition signature
\end{itemize}

\begin{figure}[h]
\centering
\includegraphics[width=0.55\textwidth]{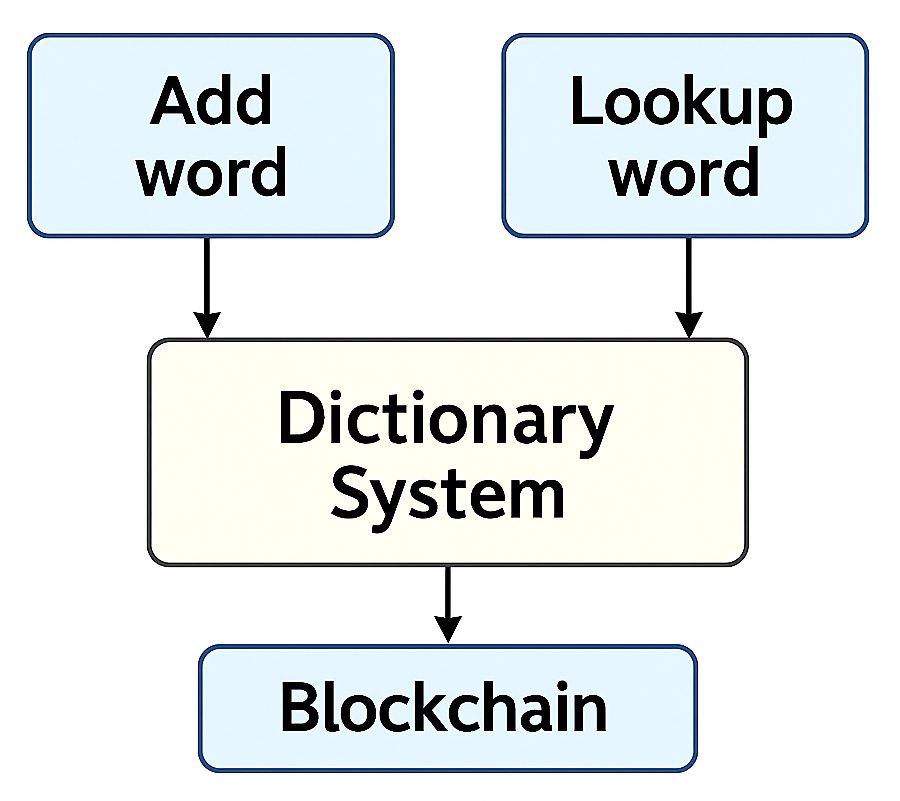}
\label{fig:blockchain-system}
\caption{PDE Dictionary System and Blockchain Structure}
\label{fig:dictionary-blockchain}
\end{figure}

\noindent
Figure~\ref{fig:dictionary-blockchain} shows how PDE's dictionary system interacts with blockchain infrastructure to ensure semantic immutability and verifiability.

\noindent
These elements are stored as individual transactions on a blockchain, guaranteeing that definitions cannot be altered retroactively. This creates a semantic consensus layer, allowing PDE to be reliably decoded across time and cultures.

\subsection{Dictionary Protocols: Registration and Retrieval}
The PDE blockchain dictionary supports the following core operations:

\medskip
\noindent
\textbf{AddDefinition}
\begin{itemize}
  \item Inputs: code, meaning, category, language, author
  \item Process: Compute SHA-256 hash → Commit to blockchain
  \item Outputs: Definition ID, Timestamp, Hash signature
\end{itemize}

\noindent
\textbf{GetDefinition}
\begin{itemize}
  \item Input: code
  \item Output: Meaning, Category, Author, Timestamp, Block ID
\end{itemize}

\noindent
\textbf{GetDictionaryVersion}
\begin{itemize}
  \item Maintains snapshots for core versions (e.g., PDE Core 100)
  \item Ensures consistency for expansion engines and human readers
\end{itemize}

\noindent
\textbf{ForkedVariants}
\begin{itemize}
  \item Supports domain-specific forks (e.g., medicine, education)
  \item Each fork includes parent ID and rationale for divergence
\end{itemize}

\noindent
This flexible architecture allows PDE dictionaries to evolve while preserving semantic fidelity and lineage.

\subsection{Toward a Global Semantic Ledger}
By grounding code definitions in a decentralized ledger, PDE creates more than just a technical reference—it forms a global semantic infrastructure.
\noindent
Future extensions may include:

\begin{itemize}
  \item A distributed dictionary governance network
  \item Cross-dictionary compatibility layers for multilingual use
  \item API-based access for PDE-to-text and text-to-PDE services
  \item Semantic audit tools for checking consistency across forks
\end{itemize}

\noindent
In this way, PDE can function as a "semantic blockchain"—a public, verifiable, and distributed knowledge ledger that safeguards the meanings of the words we encode for generations to come.

\section{Use Cases and Applications}

Permanent Data Encoding (PDE) is more than a theoretical model of symbolic compression—it is a practical framework for encoding, preserving, and transferring human knowledge in visually accessible form. This chapter explores potential applications of PDE across several domains where long-term resilience, linguistic neutrality, and technological independence are essential.

\subsection{Disaster Resilience and Post-Civilizational Recovery}
Throughout history, natural disasters, wars, and societal collapses have led to the loss of critical knowledge. In the modern era, digital records are especially fragile due to their reliance on electricity, storage media, and software infrastructure.
\noindent
PDE provides an alternative by enabling human-readable, paper-based preservation that can survive infrastructural breakdowns. Its features include:

\begin{itemize}
  \item Encoding on physical media such as paper, wood, or metal
  \item Visual interpretation without devices or electricity
  \item Reconstructability from a printed dictionary and syntax rules
  \item Cultural and linguistic neutrality across generations
\end{itemize}

\noindent
These traits make PDE particularly suited for disaster preparedness, emergency medical guides, survival manuals, and long-term cultural memory.

\subsection{Education, Literacy, and Printed Materials}
PDE can also serve as a learning tool in both monolingual and multilingual education settings:

\begin{itemize}
  \item Language learning: Learners can decode symbolic code strings into natural language sentences.
  \item Reading comprehension: Students analyze PDE-coded texts to understand structure and meaning.
  \item Printed worksheets: Teachers can provide PDE-coded materials without the need for digital devices.
  \item Syntax awareness: Learners grasp grammar and logic via structured code patterns.
\end{itemize}

\noindent
Because PDE is language-independent in structure but translatable in meaning, it can support cross-cultural education, inclusive literacy programs, and formal language instruction.

\subsection{Integration with AI and Natural Language Processing}
PDE offers a compact, logic-driven symbolic layer that integrates well with AI and language technologies:

\begin{itemize}
  \item PDE-to-Text Generation: AI models expand PDE sequences into fluent text.
  \item Text-to-PDE Encoding: NLP tools convert text into PDE strings for compression and analysis.
  \item Ambiguity Detection: PDE helps highlight vague or overloaded expressions.
  \item Semantic Compression: PDE can serve as a base for summarization and translation engines.
\end{itemize}

\noindent
This makes PDE a potential intermediate representation layer for AI communication—especially in low-resource environments.

\subsection{PDE as a Global Visual Language}
Perhaps PDE's most far-reaching potential lies in its role as a global visual language—a system that is:
\begin{itemize}
  \item Visually interpretable without reliance on sound or devices
  \item Culturally neutral yet semantically expressive
  \item Trustworthy through blockchain-based dictionary validation
  \item Flexible enough for use in education, signage, archiving, and digital interaction
\end{itemize}

\begin{figure}[h]
\centering
\includegraphics[width=0.65\textwidth]{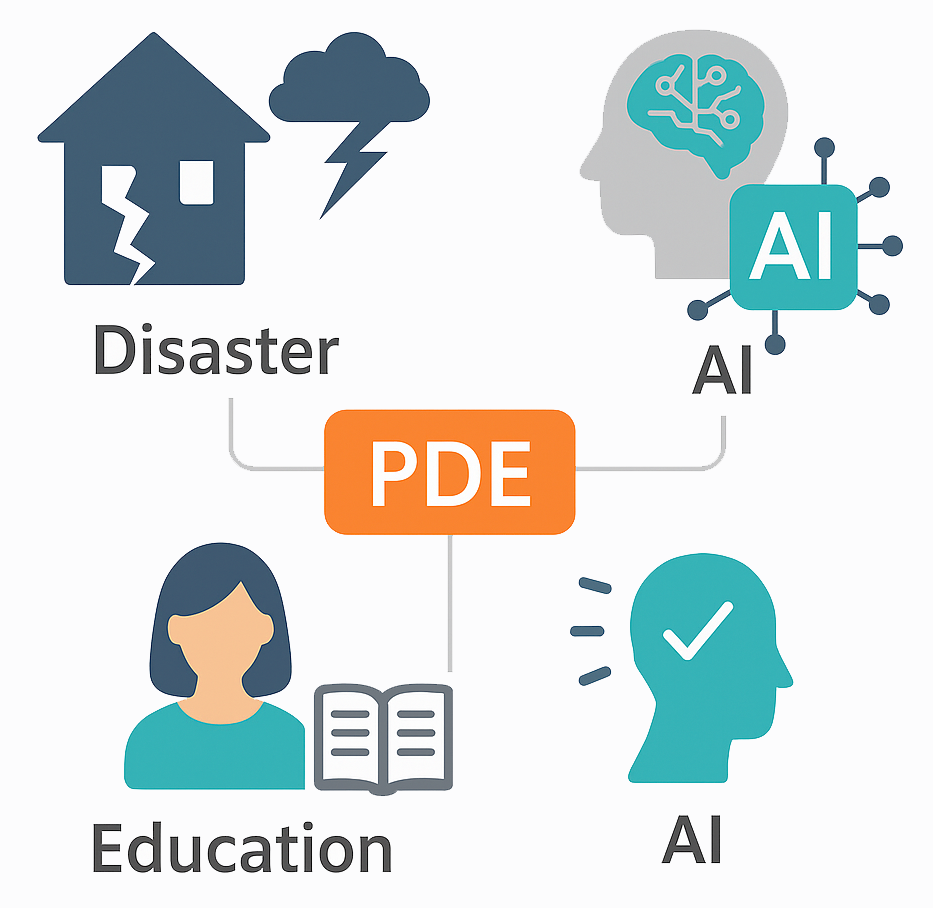}
\label{fig:pde-applications}
\caption{Use Cases and Applications of PDE Across Domains}
\label{fig:pde-usecases}
\end{figure}

\noindent
Figure~\ref{fig:pde-usecases} illustrates representative application domains of PDE in real-world contexts.

\noindent
PDE builds on the legacy of ISOTYPE, Esperanto, and Unicode, yet offers a hybrid: a human-readable, structurally rigorous, and blockchain-anchored language for the post-digital era.
In this sense, PDE is not only a visual system for encoding knowledge—it is a symbolic foundation for resilient, inclusive, and future-oriented communication.
As shown in Figure~\ref{fig:pde-usecases}, its potential reaches far beyond the boundaries of language, technology, and culture.

\subsection{Preserving Meaning Beyond Data}
As a visual-semantic interface, PDE bridges human cognition and symbolic systems.
It can operate without electricity, speak across languages, and endure the failure of machines.
In times of disaster, in classrooms, and in our interaction with intelligent systems, PDE invites us to imagine not just how we store data, but how we preserve meaning.

\section{Discussion and Limitations}

While Permanent Data Encoding (PDE) offers a novel framework for long-term, human-readable knowledge preservation, it is not without theoretical and practical challenges. This section reflects critically on the design, scalability, and adaptability of PDE, identifying key limitations and areas for future improvement.

\subsection{Compression Efficiency}
PDE assigns fixed-length 3-character codes to semantic units. Compared to binary compression algorithms (e.g., Huffman coding), PDE sacrifices space efficiency in favor of visual clarity and semantic transparency.

\begin{itemize}
  \item PDE prioritizes meaning retention over raw compression ratios.
  \item The same word may require different codes depending on context.
  \item Non-verbal data (images, audio, etc.) are not directly supported and would require additional encoding schemes.
\end{itemize}

\noindent
In this respect, PDE is better suited for robust preservation and interpretation, rather than for bandwidth-sensitive transmission.

\subsection{Dictionary Management Challenges}
The semantic dictionary is central to PDE's interpretability. However, it presents operational difficulties:

\begin{itemize}
  \item Code conflicts across independently developed dictionaries
  \item Version control issues during updates or expansions
  \item Semantic drift when definitions evolve over time
  \item Labor-intensive curation of entries and categories
\end{itemize}

\noindent
Blockchain technology partially addresses these concerns through immutability and history tracking, but international standardization and collaborative governance models are still needed.

\subsection{Syntax Expansion and Language Diversity}
PDE's syntax is currently optimized for English- and Japanese-style sentence structures. Expanding to other language families (e.g., Arabic, Hindi, Zulu) introduces complexity:

\begin{itemize}
  \item Different word orders, inflections, and grammatical markers
  \item Lack of explicit particles in some languages
  \item Varying treatment of tense, aspect, and politeness
\end{itemize}

\noindent
A truly universal system will require language-independent grammar layers, or culture-specific extensions that remain interoperable.

\subsection{Code System Complexity}
As PDE expands in expressive power, its control code set will inevitably grow. This creates risks of:

\begin{itemize}
  \item Steep learning curves for new users
  \item Increased parsing overhead for software implementations
  \item Code conflicts or ambiguity in extended sequences
\end{itemize}

\noindent
A tiered structure—e.g., PDE Core (Basic Subset) and PDE Advanced (Extended Syntax)—may help balance usability and capability across use cases.

\subsection{Philosophical Considerations}
At its core, PDE prompts a deeper question:

\noindent
Can meaning be preserved through symbolic compression?
This challenge is not merely technical—it touches on semiotics, memory, and cultural continuity.

\noindent
PDE proposes that by focusing on visually stable, structurally transparent, and semantically controlled symbols, we can build a language that outlives its own medium.

\noindent
Yet, as with any system of signs, interpretation depends on context, and languages are living systems. PDE's long-term viability will rely not just on code design, but on human commitment to preserving meaning.

\section{Conclusion and Future Directions}

Permanent Data Encoding (PDE) has been presented as a new symbolic system for compressing and preserving human knowledge in a visually readable, electrically independent, and linguistically neutral format.

\noindent
The preceding sections outlined its conceptual framework, code structure, expansion grammar, blockchain-based dictionary management, and applications across disaster resilience, education, and AI integration.

\subsection{Contributions Across Domains}
Beyond its technical innovations, PDE carries philosophical, social, and educational implications:

\begin{itemize}
  \item Technologically, it presents a new format for knowledge modeling, offering an alternative to both natural language and binary code.
  \item Socially, it offers a tool for preserving knowledge during crises, enabling education and cross-cultural understanding.
  \item Philosophically, it challenges how we define meaning, symbol, and memory across time.
\end{itemize}

\noindent
By doing so, PDE contributes to a broader rethinking of what it means to encode human understanding.

\subsection{Future Development}
To advance PDE from concept to practice, we envision the following development areas:

\begin{itemize}
  \item International standardization: Define core dictionaries, syntax specifications, and naming conventions for global adoption.
  \item API and toolkit release: Develop libraries for PDE-to-Text and Text-to-PDE conversions to facilitate integration into existing systems.
  \item Educational software: Create GUI tools for teachers and learners to construct, decode, and explore PDE codes.
  \item Open-source platform: Launch a public repository (e.g., GitHub) for collaborative dictionary building and engine development.
\end{itemize}

\noindent
We also anticipate challenges such as dictionary scalability, code proliferation, and cross-linguistic representation, all of which must be addressed through iterative refinement and community involvement.

\subsection{Toward a Visual-Semantic Infrastructure for Humanity}
In an age of increasing technological dependency and fragility, PDE proposes a humble yet radical idea:

\noindent
that meaning can be preserved in a form that is simple, visual, resilient, and shared.

\noindent
PDE is not a replacement for language, but a complementary scaffold—a minimal, symbolic substrate that encodes thought in a way that survives systems, languages, and eras.

\noindent
In this sense, PDE is both a tool for data preservation and a mirror for human thought.
Its development is not just a technical journey, but a cultural one—inviting us to reflect on what we choose to remember, and how we choose to carry it forward.

\section*{Acknowledgements}
\addcontentsline{toc}{section}{Acknowledgements}

The author would like to express deep gratitude to the many individuals who provided insights, encouragement, and constructive feedback throughout the development of the Permanent Data Encoding (PDE) framework.

\noindent
Special thanks are extended to colleagues and mentors who resonated with the PDE vision and offered valuable perspectives on its technical and philosophical foundations. Their support played a pivotal role in transforming this concept into a structured system of symbolic knowledge representation.

\noindent
The author is also grateful to the professionals who offered guidance on potential patent strategies and implementation challenges. Their candid assessments helped refine PDE into a more robust and applicable model.

\noindent
This manuscript was created with the assistance of generative AI tools, which contributed to the organization and articulation of certain sections. However, all concepts, arguments, and editorial responsibility rest solely with the author. All diagrams were drafted and finalized under the author's supervision, with AI playing a supplementary visual role.

\noindent
Above all, this work is dedicated to the memory of those who entrusted the author with a legacy of knowledge, language, and responsibility. 

\noindent
PDE is intended as a gesture of continuity, serving as a symbolic bridge between the past and the future.
\addcontentsline{toc}{section}{Acknowledgements}

\section*{References}
\begin{enumerate}
    \item Nakamoto, S. (2008). \textit{Bitcoin: A Peer-to-Peer Electronic Cash System}.
    \item W3C (2023). \textit{Decentralized Identifiers (DIDs) v1.0}. \url{https://www.w3.org/TR/did-core/}
    \item Eco, U. (1976). \textit{A Theory of Semiotics}. Indiana University Press.
    \item ISO/IEC 10646:2020. \textit{Universal Coded Character Set (UCS)}.
    \item Marr, D. (1982). \textit{Vision: A Computational Investigation into the Human Representation and Processing of Visual Information}. MIT Press.
    \item Bostrom, N. (2003). The Future of Human Evolution. \textit{Journal of Evolution and Technology}, Vol. 9.
    \item Crystal, D. (2003). \textit{English as a Global Language}. Cambridge University Press.
    \item OpenAI (2023). \textit{Technical Report on GPT-4}. \url{https://openai.com/research/gpt-4}
\end{enumerate}

\end{document}